\documentclass[11pt,a4paper]{report}

\usepackage{amssymb}
\usepackage{amsmath}
\usepackage{amsfonts}
\usepackage{siunitx}
\usepackage{graphicx}
\usepackage{booktabs}
\usepackage{titlesec}
\usepackage{geometry}
\usepackage{xcolor}
\usepackage[HWP]{dhucs-interword}
\usepackage{enumitem}
\usepackage{float}
\usepackage{setspace}
\usepackage{etoolbox} 

\usepackage[pdfauthor={현재익},
            pdftitle={Log-Time K-Means Clustering for 1D Data},
            pdfsubject={Log-Time K-Means Clustering for 1D Data: Novel Approaches with Proof and Implementation},
            pdfkeywords={서울대학교, 공과대학, 컴퓨터공학부, 졸업논문, SNU, ENG, CSE, thesis, k-means},
            ]{hyperref}

\titleformat{\section} 
  {\normalfont\fontsize{14pt}{14pt}\bfseries} 
  {} 
  {0pt} 
  {\thesection\quad} 

\titleformat{\chapter}[block]
  {\centering\normalfont\fontsize{16pt}{16pt}\bfseries} 
  {\thechapter.}                       
  {10pt}                               
  {}

\titlespacing*{\chapter}
  {0pt}    
  {-30pt}  
  {20pt}   

\usepackage{listings}
\begin{document}

\pagestyle{empty} 
\begin{center}
    \vspace*{-1.0cm}
    \fontsize{24pt}{30pt}\selectfont Log-Time K-Means for 1D Data: \\
    \fontsize{20pt}{24pt}\selectfont Novel Approaches with Proof and Implementation \\
    \vspace{0.5cm}
    \fontsize{18pt}{22pt}\selectfont (1D 데이터에 대한 로그 시간 K-평균 클러스터링: \\
    새로운 접근법의 증명과 구현) \\
    \vspace{2cm}
    \fontsize{18pt}{18pt}\selectfont
    지도교수: 허 충 길 \\
    \vspace{2cm}
    \fontsize{20pt}{20pt}\selectfont
    이 논문을 공학학사 학위 논문으로 제출함. \\
    \vspace{2cm}
    \fontsize{18pt}{18pt}\selectfont
    2024년~~~12월~~~19일 \\
    \vfill
    \fontsize{20pt}{20pt}\selectfont
    서울대학교 공과대학 \\
    컴~~퓨~~터~~공~~학~~부 \\
    현~~재~~익 \\
    \vspace{2cm}
    \fontsize{20pt}{20pt}\selectfont
    2025년~~~2월 \\
\end{center}
\clearpage

\pagestyle{plain}

\pagenumbering{Roman}
\setcounter{page}{1} 
\chapter*{Abstract}
\addcontentsline{toc}{chapter}{Abstract}
\vspace*{-1.1cm}
{
\centering
\fontsize{22pt}{22pt}\selectfont Log-Time K-Means for 1D Data: \\
\fontsize{18pt}{20pt}\selectfont Novel Approaches with Proof and Implementation \\
}

\vspace*{0.8cm}

{
\flushright
\fontsize{14pt}{16pt}\selectfont
Jake Hyun \\
Computer Science and Engineering \\ 
College of Engineering \\
Seoul National University \\
}
\vspace*{0.8cm}

Clustering is a key task in machine learning, with $k$-means being widely used for its simplicity and effectiveness. While 1D clustering is common, existing methods often fail to exploit the structure of 1D data, leading to inefficiencies. This thesis introduces optimized algorithms for $k$-means++ initialization and Lloyd's algorithm, leveraging sorted data, prefix sums, and binary search for improved computational performance.

The main contributions are: (1) an optimized \(k\)-cluster algorithm achieving \(O(l \cdot k^2 \cdot \log n)\) complexity for greedy $k$-means++ initialization and \(O(i \cdot k \cdot \log n)\) for Lloyd's algorithm, where \(l\) is the number of greedy $k$-means++ local trials, and \(i\) is the number of Lloyd's algorithm iterations, and (2) a binary search-based two-cluster algorithm, achieving \(O(\log n)\) runtime with deterministic convergence to a Lloyd’s algorithm local minimum.

Benchmarks demonstrate over a 4500x speedup compared to \texttt{scikit-learn} for large datasets while maintaining clustering quality measured by within-cluster sum of squares (WCSS). Additionally, the algorithms achieve a 300x speedup in an LLM quantization task, highlighting their utility in emerging applications. This thesis bridges theory and practice for 1D $k$-means clustering, delivering efficient and sound algorithms implemented in a JIT-optimized open-source Python library.

\vfill
\textbf{Keywords: $k$-means clustering, Lloyd’s algorithm, $k$-means++ initialization, one-dimensional clustering, binary search, prefix sums}

\clearpage

\renewcommand\baselinestretch{1.3}
\tableofcontents
\cleardoublepage

\pagenumbering{arabic} 
\setcounter{page}{1}   
\chapter{Introduction}\label{chap:introduction}

Clustering is a fundamental task in data analysis and machine learning, with applications in diverse fields such as image segmentation, natural language processing, financial modeling, and bioinformatics \cite{Xu2005Survey}. Among clustering methods, $k$-means \cite{MacQueen1967} is one of the most widely used algorithms due to its conceptual simplicity and computational efficiency. However, finding the optimal solution to the $k$-means problem is NP-hard in general for \(d\)-dimensional data \cite{nphard}, prompting practical implementations to rely on heuristic approaches such as Lloyd’s algorithm \cite{lloyd,max}.

One-dimensional (1D) clustering problems arise frequently in a wide range of real-world scenarios, including social network analysis, bioinformatics, and the retail market \cite{Arnaboldi_2016,genomemining,retail}. For this special case, there have been significant advances in achieving globally optimal solutions efficiently. Wang and Song \cite{wang1ddp} introduced a \(O(k \cdot n^2)\) dynamic programming algorithm for the 1D $k$-means problem, and Grønlund et al. \cite{fastexactkmeans} later improved this to \(O(n)\), demonstrating that optimal clustering can be computed in linear time for one dimension.

While globally optimal algorithms for the 1D $k$-means problem exist, they are not always suitable for scenarios where speed and scalability are paramount. In many real-world applications—particularly those involving large datasets or latency-critical tasks—achieving a near-optimal solution quickly can be more valuable than computing the exact global minimum. Practical libraries, such as \texttt{scikit-learn}'s $k$-means \cite{sklearn}, do not exploit the structure of 1D data and instead treat it as a general case, leaving room for further optimization. Under such conditions, improving Lloyd’s algorithm for 1D data provides a route to faster performance.

This thesis presents a novel set of algorithms that optimize Lloyd’s algorithm for the 1D setting. By carefully exploiting the properties of sorted data, these methods achieve logarithmic runtime, dramatically reducing computational costs while maintaining high-quality clustering outcomes. The contributions include the following:

\begin{enumerate}[leftmargin=*]
\item \textbf{An optimized $k$-means++ initialization and Lloyd’s algorithm} for approximating general \(k\)-cluster problems in one dimension. By carefully leveraging the properties of sorted data, the proposed approach replaces the linear dependence on the dataset size \(n\) with logarithmic factors, resulting in substantial runtime improvements. Specifically, the greedy $k$-means++ initialization achieves a time complexity of \(O(l \cdot k^2 \cdot \log n)\), where \(l\) is the number of local trials, followed by Lloyd’s algorithm iterations with \(O(i \cdot k \cdot \log n)\), where $i$ is the number of iterations. Additional preprocessing, such as sorting and prefix sum calculations, contributes \(O(n \log n)\) and \(O(n)\), respectively, when required.

This method improves upon standard $k$-means implementations, where greedy $k$-means++ initialization requires \(O(l \cdot k \cdot n)\) time, and Lloyd's algorithm iterations require \(O(i \cdot k \cdot n)\). By reducing the dependence on \(n\), the dataset size, the proposed optimizations achieve significant speedups, as experimentally demonstrated in Section~\ref{sec:runtime_performance}.

\item \textbf{A binary search-based algorithm for the two-cluster case}, which achieves \(O(\log n)\) runtime and deterministically converges to a Lloyd’s algorithm solution, skipping iterative refinements entirely. Additional preprocessing costs include \(O(n)\) for prefix sums and \(O(n \log n)\) for sorting, if not already provided. While the global minimum is not guaranteed, this method is highly desirable for scenarios requiring very fast and deterministic clustering.
\end{enumerate}

Along with the thesis, a complete library implementation in Python 3 is provided, optimized with Numba just-in-time (JIT) compilation \cite{numba} to enable efficient integration into various applications.

Benchmarks against the highly optimized and widely used \texttt{scikit-learn} $k$-means implementation \cite{sklearn} highlight the efficiency of these algorithms, as detailed in Section~\ref{sec:runtime_performance}. The results demonstrate the following:

\begin{itemize}[leftmargin=*]
    \item \textbf{Orders-of-magnitude speedups}, even when including preprocessing steps such as sorting and prefix sum calculations.
    \item \textbf{Equivalent or comparable clustering results} in terms of within-cluster sum of squares (WCSS), the objective function of K-means.
\end{itemize}

The proposed algorithms also find utility in emerging and highly relevant applications such as \textbf{quantization for large language models (LLMs)}, where efficient quantization can be achieved by running 1D $k$-means clustering on model weights \cite{sqllm}. In particular, cutting-edge quantization methods like Any-Precision LLM \cite{anyprec} rely on repeated executions of the two-cluster approach to hierarchically subdivide clusters of weights. The novel algorithm presented here is exceptionally well-suited for such scenarios, providing over a \textbf{300-fold speedup} compared to \texttt{scikit-learn}, as demonstrated in Section~\ref{sec:llm_quantization}. This practical importance is underscored by the direct use of the proposed library implementation, \texttt{flash1dkmeans}, within the official Any-Precision LLM implementation\footnote{\url{https://github.com/SNU-ARC/any-precision-llm}}.

Overall, this thesis contributes both theoretical advancements and practical tools for one-dimensional $k$-means clustering. By providing a rigorous theoretical foundation and an optimized implementation, the proposed methods demonstrate the feasibility and efficiency of adapting $k$-means and Lloyd’s algorithm to the one-dimensional setting. These contributions establish not only a proof of concept but also a practical solution ready for deployment in diverse computational tasks.

\chapter{Background}\label{chap:background}

This chapter provides the necessary background on the $k$-means clustering problem, Lloyd's algorithm, and the $k$-means++ initialization, along with an overview of relevant works. The focus is on the theoretical foundations, time complexities, and practical implementations relevant to this thesis.

\section{$k$-Means Clustering}\label{sec:kmeans}

The $k$-means clustering problem is a widely studied unsupervised learning problem. Given a set of \(n\) data points \(X = \{x_1, x_2, \dots, x_n\}\) in a metric space and a positive integer \(k\), the goal of \(k\)-means clustering is to partition the data into \(k\) disjoint clusters \(C_1, C_2, \dots, C_k\) such that the within-cluster sum of squared distances (WCSS) is minimized. Formally, the objective function is:

\[
\text{WCSS} = \sum_{i=1}^k \sum_{x \in C_i} \|x - \mu_i\|^2,
\]

where \(\mu_i\) is the centroid of cluster \(C_i\), defined as the mean of all points in \(C_i\). 

Finding the globally optimal solution to the $k$-means problem is NP-hard in general for \(d\)-dimensional data, even for \(k = 2\) \cite{nphard}. As a result, heuristic algorithms such as Lloyd's algorithm are commonly used in practice to approximate solutions efficiently.

\section{Lloyd’s Algorithm}\label{sec:lloyds}

Lloyd’s algorithm \cite{lloyd,max} is a popular iterative method for solving the $k$-means problem. It alternates between assigning data points to their nearest cluster centroid and updating the centroids based on the current cluster assignments. The steps of the algorithm are as follows:

\begin{enumerate}[leftmargin=*]
    \item \textbf{Initialization:} Choose \(k\) initial centroids \(\mu_1, \mu_2, \dots, \mu_k\).
    \item \textbf{Assignment step:} Assign each point \(x_j \in X\) to the cluster \(C_i\) with the closest centroid:
    \[
    C_i = \{x_j \mid \|x_j - \mu_i\| \leq \|x_j - \mu_m\|, \, \forall m \neq i\}.
    \]
    \item \textbf{Update step:} Update each centroid \(\mu_i\) as the mean of all points assigned to \(C_i\):
    \[
    \mu_i = \frac{1}{|C_i|} \sum_{x \in C_i} x.
    \]
    \item Repeat the assignment and update steps until convergence, typically when the centroids no longer change significantly or a maximum number of iterations is reached.
\end{enumerate}

\noindent \textbf{Remark:}  
Both the assignment and update steps of Lloyd's algorithm ensure that the within-cluster sum of squares (WCSS) decreases monotonically after each iteration. As a result, the algorithm is guaranteed to converge to a local minimum of the WCSS objective function.

\noindent \textbf{Time Complexity:}  
The time complexity of Lloyd’s algorithm for general \(d\)-dimensional data is \(O(i \cdot k \cdot n \cdot d)\), where \(i\) is the number of iterations. For 1D data, the complexity simplifies to \(O(i \cdot k \cdot n)\).

\section{$k$-Means++ Initialization}\label{sec:kmeanspp}

The quality of solutions obtained by Lloyd’s algorithm depends heavily on the choice of initial centroids. The $k$-means++ initialization algorithm \cite{kmeansplusplus} improves the centroid selection process by probabilistically choosing points based on their distances to already selected centroids. The steps of the algorithm are as follows:

\begin{enumerate}[leftmargin=*]
    \item \textbf{Initialization:} Choose the first centroid \(\mu_1\) uniformly at random from the data points.
    \item \textbf{Candidate selection:} For each subsequent centroid \(\mu_i\), select a point \(x_j\) with probability proportional to its squared distance from the nearest already chosen centroid:
    \[
    P(x_j) = \frac{\text{Distance}(x_j, C_{\text{nearest}})^2}{\sum_{x_i \in X} \text{Distance}(x_i, C_{\text{nearest}})^2}.
    \]
    \item Repeat the candidate selection step until \(k\) centroids are chosen.
\end{enumerate}

\noindent \textbf{Remark:}  
The $k$-means++ initialization improves the spread of centroids compared to random initialization, significantly reducing the likelihood of poor clustering outcomes. However, this improvement comes at the cost of additional computation during the initialization phase.

\noindent \textbf{Time Complexity:}  
The time complexity of standard $k$-means++ initialization for \(d\)-dimensional data is \(O(k \cdot n \cdot d)\), where \(k\) is the number of clusters, \(n\) is the number of data points, and \(d\) is the dimensionality of the data.

\noindent \textbf{Greedy $k$-Means++ Initialization:}  
In the greedy version of $k$-means++ initialization, briefly mentioned in the conclusion of the original paper \cite{kmeansplusplus}, \(l\) candidate centroids are evaluated at each step, and the one minimizing the WCSS is selected. The total time complexity for the greedy version is \(O(l \cdot k \cdot n \cdot d)\), where \(l\) is the number of local trials. A common choice for \(l\) is \(O(\log k)\) \cite{localtrials1, localtrials2}, and \texttt{scikit-learn} adopts a similar approach with \(l = 2 + \log k\) \cite{sklearn}. For 1D data, where \(d = 1\), the complexity simplifies to \(O(l \cdot k \cdot n)\).

\section{Weighted $k$-Means}\label{sec:weightedkmeans}

Weighted $k$-means generalizes the standard $k$-means problem by assigning each data point \(x_j\) a weight \(w_j\). The objective function becomes:

\[
\text{WCSS} = \sum_{i=1}^k \sum_{x_j \in C_i} w_j \|x_j - \mu_i\|^2,
\]

where the centroid \(\mu_i\) is updated as the weighted mean:

\[
\mu_i = \frac{\sum_{x_j \in C_i} w_j \cdot x_j}{\sum_{x_j \in C_i} w_j}.
\]

\noindent \textbf{Changes to Lloyd’s Algorithm:}  
The \textbf{update step} computes weighted centroids instead of simple means. The assignment step remains unchanged.

\noindent \textbf{Changes to $k$-Means++ Initialization:}  
The probability of selecting a point \(x_j\) as a centroid becomes proportional to its weighted squared distance from the nearest already chosen centroid:

\[
P(x_j) = \frac{w_j \cdot \text{Distance}(x_j, C_{\text{nearest}})^2}{\sum_{x_i \in X} w_i \cdot \text{Distance}(x_i, C_{\text{nearest}})^2}.
\]

\noindent \textbf{Relevance:}  
Weighted $k$-means is particularly useful in applications where data points contribute unequally to the clustering objective, such as quantization for large language models (LLMs) \cite{sqllm, anyprec}.

\noindent \textbf{Implementation Note:}  
The algorithms detailed in this thesis support both unweighted and weighted $k$-means clustering, ensuring flexibility for practical use cases.

\section{Relevant Works}\label{sec:relevantworks}

For the special case of 1D $k$-means clustering, significant progress has been made in achieving globally optimal solutions. Wang and Song \cite{wang1ddp} introduced a dynamic programming algorithm with a time complexity of \(O(k \cdot n^2)\). This was later improved by Grønlund et al. \cite{fastexactkmeans}, who developed an \(O(n)\)-time algorithm for computing the exact optimal clustering in 1D.

More recently, Froese et al. \cite{borderkmeans} proposed the Border $k$-Means algorithm, which optimizes 1D clustering by introducing deterministic border adjustments and achieves \(O(n \log n)\) time complexity overall, dominated by the initial sorting step. The cluster update phase is \(O(n)\) in practice for most datasets but can scale up to \(O(n \log n)\) under certain conditions. While the algorithm produces deterministic results equivalent to Lloyd's algorithm, it does not guarantee globally optimal clustering solutions and focuses instead on practical efficiency.

Similarly, both Border $k$-Means and our \(k\)-cluster algorithm focus on efficient but non-optimal clustering. However, our approach differs in two key aspects. First, it directly integrates \(k\)-means++ initialization to improve cluster seeding. Second, it achieves \(O(\log n)\) time complexity for updates after an initial sorting and preprocessing phase, offering a significant improvement in efficiency for large-scale datasets.

\sloppy
In contrast to these specialized approaches, widely-used libraries such as \texttt{scikit-learn} \cite{sklearn} implement general-purpose $k$-means algorithms that are not optimized for the 1D case. \texttt{scikit-learn}'s $k$-means algorithm treats 1D data as a general instance of higher-dimensional clustering. While it leverages Cython \cite{cython} for efficient performance, it operates with \(O(l \cdot k \cdot n)\) initialization time and \(O(i \cdot k \cdot n)\) iteration time for 1D data, significantly limiting its scalability compared to the specialized methods discussed above.

This thesis addresses the gap in optimizing Lloyd’s algorithm and $k$-means++ initialization specifically for 1D clustering, targeting scenarios where speed and scalability are paramount. By combining \(k\)-means++ initialization with logarithmic-time updates, our approach achieves a balance between computational efficiency and clustering quality, offering a practical solution for real-world applications.
\chapter{Novel Approaches and Proof of Validity}\label{chap:algorithms}

This chapter presents our proposed approaches for solving general \(k\)-cluster problems in one dimension. We introduce two algorithms: the \textit{\(k\)-cluster algorithm} and the \textit{2-cluster algorithm}. Both methods exploit the structure of one-dimensional data and utilize sorting, prefix sums, and binary search to achieve significant computational efficiency compared to traditional clustering methods.

\paragraph{Finding Cluster Boundaries:}
For one-dimensional data, determining a point's cluster assignment involves identifying the interval it falls into. The boundaries between clusters are the arithmetic midpoints of consecutive centroids. When both the data and centroids are sorted, these boundaries can be efficiently located using binary search, requiring $O(k \cdot \log n)$ time. If centroids need sorting, an additional $O(k \cdot \log k)$ time is needed; however, as $k \leq n$, the total time remains $O(k \cdot \log n)$. This approach forms the basis for subsequent optimizations.

\section{The \(k\)-Cluster Algorithm}\label{sec:k_cluster_algorithm}

The \(k\)-cluster algorithm is a one-dimensional adaptation of greedy $k$-means++ initialization followed by Lloyd’s algorithm iterations, and can be defined for both weighted and unweighted data. The algorithm leverages:

\begin{enumerate}
    \item \textbf{Sorted Data:} Sorting the input array \(X\) of size \(n\) in ascending order allows quick determination of cluster assignments via binary search on cluster boundaries.
    \item \textbf{Prefix Sums:} Precomputing prefix sums enables constant-time computation of weighted and unweighted sums, means, and inertia values over arbitrary intervals. Specifically, for weighted data:
    \[
    W[j] = \sum_{i=1}^j w_i, \quad 
    (WX)[j] = \sum_{i=1}^j w_i x_i, \quad 
    (WX^2)[j] = \sum_{i=1}^j w_i x_i^2.
    \]
    For unweighted data, this simplifies to:
    \[
    X^{(1)}[j] = \sum_{i=1}^j x_i, \quad X^{(2)}[j] = \sum_{i=1}^j x_i^2.
    \]
    \item \textbf{Binary Search:} Binary search is central to efficiently determining cluster boundaries and performing weighted random sampling during initialization.
\end{enumerate}

\subsection{WCSS and Prefix Sums}

The within-cluster sum of squares (WCSS) for weighted data is defined as:
\[
\text{WCSS} = \sum_{i=1}^k \sum_{x_j \in C_i} w_j (x_j - \mu_i)^2,
\]
where \(C_i\) is the \(i\)-th cluster, \(\mu_i\) is its centroid, and \(w_j\) is the weight of point \(x_j\). Expanding the squared term:
\[
(x_j - \mu_i)^2 = x_j^2 - 2x_j\mu_i + \mu_i^2,
\]
yields:
\[
\text{WCSS}_i = \sum_{x_j \in C_i} w_j x_j^2 - 2\mu_i \sum_{x_j \in C_i} w_j x_j + \mu_i^2 \sum_{x_j \in C_i} w_j.
\]

Using prefix sums:
\[
\begin{aligned}
\sum_{x_j \in C_i} w_j x_j^2 &= (WX^2)[b_i] - (WX^2)[b_{i-1}], \\
\sum_{x_j \in C_i} w_j x_j &= (WX)[b_i] - (WX)[b_{i-1}], \\
\sum_{x_j \in C_i} w_j &= W[b_i] - W[b_{i-1}].
\end{aligned}
\]

where \(b_{i-1}\) and \(b_i\) are the cluster boundaries. The centroid is:
\[
\mu_i = \frac{(WX)[b_i] - (WX)[b_{i-1}]}{W[b_i] - W[b_{i-1}]}.
\]

All these queries take \(O(1)\) time per cluster once the prefix sums are computed. Hence, WCSS and centroid calculations are efficient, requiring only \(O(k)\) time across all $k$ clusters, if cluster boundaries are known. Determining cluster boundaries costs \(O(k \log n)\), so the total cost for WCSS calculation given centroids is \(O(k \log n)\).

\subsection{Greedy $k$-Means++ Initialization}

The $k$-means++ initialization selects centroids such that new centroids are chosen with probabilities proportional to their squared distances from the closest existing centroid. Our method efficiently implements this using \textbf{binary search} combined with \textbf{cumulative sum queries}.

\paragraph{Steps for Initialization:}
\begin{enumerate}
    \item \textbf{First Centroid Selection:}
    The first centroid is chosen randomly, weighted by the point weights \(w_j\). To achieve this:
    \begin{itemize}
        \item Given the cumulative sum of weights \(W[j] = \sum_{i=1}^j w_i\),
        \item Generate a random number \(r \in [0, W[n]]\),
        \item Perform binary search on \(W\) to find the smallest $j$ such that $W[j] \geq r$, and thus the corresponding point \(x_j\). This step costs \(O(\log n)\).
    \end{itemize}

    \item \textbf{Subsequent Centroid Selection:}
    For each new centroid:
    \begin{enumerate}
        \item \textbf{Binary Search for Cluster Assignments:}
        Given the existing centroids, determine the cluster boundaries using a binary search of consecutive centroid midpoints. This step costs \(O(k \log n)\).

        \item \textbf{Cumulative Sum for Squared Distances:}
        To sample a new centroid, we need the cumulative sum of squared distances \(D_i^2\), where \(D_i\) is the distance of \(x_i\) to its closest centroid. The cumulative sum \(S[j]\) is defined as:
        \[
        S[j] = \sum_{i=1}^j D_i^2.
        \]
        \textbf{Importantly:} \(S\) is not explicitly constructed. Instead for each query $j$ on $S$:
        \begin{itemize}
            \item The sum of squared distances \(D_i^2\) are obtained using prefix sums over the \(k\) clusters, up to the $j$th point. This is equivalent to calculating the WCSS up to the $j$th point. For each cluster, this sum can be retrieved in $O(1)$ time, as the cluster boundaries are known. 
            \item Querying \(S[j]\) for any \(j\) requires \(O(k)\) time, as it aggregates contributions from all relevant clusters.
        \end{itemize}

        \item \textbf{Binary Search on \(S\):}
        \begin{itemize}
            \item Generate a random number \(r \in [0, S[n]]\),
            \item Perform binary search on \(S\) to find the smallest \(j\) such that \(S[j] \geq r\).
        \end{itemize}
        Each binary search involves \(O(\log n)\) queries of \(S\), where each query takes \(O(k)\). Thus, the total cost for sampling one new centroid is \(O(k \cdot \log n)\), and for $l$ candidates, $O(l \cdot k \cdot \log n)$.

        \item \textbf{Greedy Candidate Selection:}
        For each candidate:
        \begin{itemize}
            \item Update cluster boundaries using binary search (\(O(k \cdot \log n)\)),
            \item Compute the total WCSS using prefix sums (\(O(k)\)).
        \end{itemize}
        The candidate minimizing the total WCSS is selected as the next centroid. For \(l\) candidates, this step costs \(O(l \cdot k \cdot \log n)\).

        \item \textbf{Combined Initialization Time Complexity:}
        Combining the steps, the total cost for generating and evaluating \(l\) candidates per new centroid is $O(k \cdot \log n \cdot + l \cdot k \cdot \log n + l \cdot k \cdot \log n) = O(l \cdot k \cdot \log n)$.
    \end{enumerate}
\end{enumerate}

\subsection{Complexity Analysis}

The overall time complexity of the \(k\)-cluster algorithm is as follows:

\paragraph{Greedy $k$-Means++ Initialization:}
As detailed in the previous section:
\begin{itemize}
    \item Selecting the first centroid using weighted sampling costs \(O(\log n)\),
    \item Each subsequent centroid requires \(O(l \cdot k \cdot \log n)\), where \(l\) is the number of local trials.
\end{itemize}
The total cost for initialization across \(k\) centroids is therefore:
\[
O(l \cdot k^2 \cdot \log n).
\]

\paragraph{Lloyd’s Algorithm Iterations:}
Each iteration of Lloyd’s algorithm consists of:
\begin{itemize}
    \item Updating cluster boundaries via binary search: \(O(k \cdot \log n)\),
    \item Updating centroids using prefix sums: \(O(k)\).
\end{itemize}
For \(i\) iterations, the total cost is:
\[
O(i \cdot k \cdot \log n).
\]

\paragraph{Overall Time Complexity:}
The combined cost of greedy $k$-means++ initialization and Lloyd’s algorithm is:
\[
O(l \cdot k^2  \cdot \log n) + O(i \cdot k \cdot \log n).
\]

This does not account for the initial overhead of sorting the data and calculating prefix sums, which cost $O(n \log n)$ and $O(n)$, respectively.

Comparing against conventional implementations of $O(l \cdot k \cdot n) + O(i \cdot k \cdot n)$, note how the dependence on $n$ (dataset size) has decreased, at the cost of quadratic complexity in $k$ during initialization. However, since $k \ll n$ in most practical cases, this tradeoff is justified. For experimental speedup proofs, see Chapter~\ref{chap:experiments}.

\section{The 2-Cluster Algorithm}

For the 2-cluster problem in one-dimensional sorted data, the task reduces to finding a single \textbf{cluster boundary} that divides the data into two contiguous clusters. To efficiently locate this boundary, we iteratively refine a \textbf{search scope} to identify the correct \textbf{division interval}. A division interval is defined as the interval between two consecutive points in the sorted data that contains the cluster boundary.

\textbf{Note:} For all discussions in this section, all notions of direction (i.e., left or right) are with respect to the one-dimensional coordinate axis along which the data points are sorted. Thus, \emph{left} refers to decreasing \(x\)-values and \emph{right} refers to increasing \(x\)-values.

\subsection{Definitions and Key Observations}

\begin{itemize}
    \item A \textbf{division interval} is defined as the interval between two consecutive points \(x_{\text{div\_left}}\) and \(x_{\text{div\_right}}\) in sorted data that contains the cluster boundary (note that, of course, $\text{div\_left} + 1 = \text{div\_right}$).
    \item The \textbf{midpoint} for a division interval is defined as:
    \[
    \text{Midpoint} = \frac{\mu_{\text{left}} + \mu_{\text{right}}}{2},
    \]
    where \(\mu_{\text{left}}\) and \(\mu_{\text{right}}\) are the centroids of the left and right clusters defined by the division interval, respectively. These centroids are computed with prefix sums, using:
    \[
    \mu_{\text{left}} = \frac{\sum_{i=1}^{\text{div\_left}} w_i x_i}{\sum_{i=1}^{\text{div\_left}} w_i}, \quad 
    \mu_{\text{right}} = \frac{\sum_{i=\text{div\_right}}^n w_i x_i}{\sum_{i=\text{div\_right}}^n w_i}.
    \]
    The prefix sums $W$ and $WX$, as defined for the $k$-cluster algorithm, allow this calculation to be done in $O(1)$ time.
    \item A division interval is classified as follows:
        \begin{itemize}
            \item \textbf{Right-pointing:} The midpoint lies to the right of \(x_{\text{div\_right}}\).
            \item \textbf{Left-pointing:} The midpoint lies to the left of \(x_{\text{div\_left}}\).
            \item \textbf{Convergent:} The midpoint lies within the division interval itself, indicating a Lloyd’s algorithm convergence.
        \end{itemize}
    Note that every division interval can be classified into exactly one of these three categories.

    \item The \textbf{search scope} refers to the range of candidate division intervals, which is iteratively refined during the binary search to locate a convergent interval.
\end{itemize}

\subsection{Algorithm Description}

The algorithm aims to identify the correct division interval (i.e., a convergent interval) using binary search. The key steps are as follows:

\begin{enumerate}
    \item \textbf{Initialize the Search Scope:}  
    Start with the whole scope---that is, all possible division intervals ranging from the first interval \([x_1, x_2]\) to the last interval \([x_{n-1}, x_n]\).

    \item \textbf{Iteratively Query the Center Interval:}  
    At each step:
    \begin{itemize}
        \item Select the \textbf{center division interval} within the current search scope.
        \item Compute the centroids \(\mu_{\text{left}}\) and \(\mu_{\text{right}}\) and calculate the midpoint:
        \[
        \text{Midpoint} = \frac{\mu_{\text{left}} + \mu_{\text{right}}}{2}.
        \]
    \end{itemize}

    \item \textbf{Refine the Search Scope:}
    Compare the midpoint to the endpoints $x_{\text{div\_left}}$ and $x_{\text{div\_right}}$ of the queried division interval:
    \begin{itemize}
        \item If the interval is \textbf{right-pointing}, exclude all intervals to the left of the current interval, including itself.
        \item If the interval is \textbf{left-pointing}, exclude all intervals to the right of the current interval, including itself.
        \item If the interval is \textbf{convergent}, terminate; the cluster boundary has been found.
    \end{itemize}

    \item \textbf{Repeat Until Convergence:}  
    Continue the process until a convergent interval is found.
\end{enumerate}

The binary search guarantees that the number of candidate intervals is halved at each iteration, ensuring \(O(\log n)\) convergence.

\subsection{Proof of Validity}

To prove the correctness of the algorithm, we rely on the monotonic behavior of the centroids’ midpoint and the structure of the division intervals.

\paragraph{1. Monotonic Behavior of the Midpoint:}  
When the division interval \([x_{\text{div\_left}}, x_{\text{div\_right}}]\) is shifted one step to the right, i.e., to \([x_{\text{div\_right}}, x_{\text{div\_right} + 1}]\):
\begin{itemize}
    \item The point \(x_{\text{div\_right}}\), which was previously the leftmost point in the right cluster, is excluded from the right cluster and added to the left cluster.
    \item This change causes the \textbf{left centroid} \(\mu_{\text{left}}\) to increase, as a point with a larger value has been included in the left cluster.
    \item Simultaneously, the \textbf{right centroid} \(\mu_{\text{right}}\) also increases, as the smallest point in the right cluster has been removed.
    \item Since both centroids increase, the new \textbf{midpoint}:
    \[
    \text{Midpoint}^{\text{new}} = \frac{\mu_{\text{left}}^{\text{new}} + \mu_{\text{right}}^{\text{new}}}{2}
    \]
    is greater than the old midpoint:
    \[
    \text{Midpoint}^{\text{new}} > \text{Midpoint}^{\text{old}}.
    \]
\end{itemize}

Similarly, shifting the division interval one step to the left causes the midpoint to strictly decrease.

\paragraph{2. Behavior of Right-Pointing and Left-Pointing Intervals:}  
The monotonic behavior of the midpoint ensures the following:

\begin{itemize}
    \item \textbf{Right-Pointing Intervals:}  
    Suppose a division interval \([x_{\text{div\_left}}, x_{\text{div\_right}}]\) is right-pointing, so:
    \[
    x_{\text{div\_right}} < \text{Midpoint}^{\text{old}}.
    \]
    After shifting the interval one step to the right to the new interval \([x_{\text{div\_right}}, x_{\text{div\_right+1}}]\), the corresponding new midpoint strictly increases:
    \[
    x_{\text{div\_right}} < \text{Midpoint}^{\text{old}} < \text{Midpoint}^{\text{new}}.
    \]
    For the new interval  \([x_{\text{div\_right}}, x_{\text{div\_right+1}}]\) to become left-pointing, the new midpoint, $\text{Midpoint}^{\text{new}}$, would have to be less than \(x_{\text{div\_right}}\). But directly from the inequality above:
    \[
    x_{\text{div\_right}} < \text{Midpoint}^{\text{new}}.
    \]
    Thus, a right-pointing interval cannot suddenly become left-pointing after a rightward shift; it remains right-pointing or becomes convergent.

    \item \textbf{Left-Pointing Intervals:}  
    By a symmetric argument, for a left-pointing interval, shifting one step to the left strictly decreases the midpoint, ensuring it cannot suddenly become right-pointing. Instead, it remains left-pointing or becomes convergent.
\end{itemize}

This ensures that from the perspective of shifting the intervals (one step at a time in the chosen direction), right-pointing and left-pointing intervals cannot directly switch roles in a single move.

\paragraph{Convergent Interval in a Search Scope}
In any search scope where the first interval is right-pointing and the last interval is left-pointing, there must exist at least one convergent interval between them. This follows from the fact that a right-pointing interval cannot immediately precede a left-pointing interval when moving stepwise to the right.

\paragraph{Convergence of the Binary Search}
For the entire search scope, the first division interval (between \(x_1\) and \(x_2\)) is either right-pointing or convergent, and the last division interval (between \(x_{n-1}\) and \(x_n\)) is either left-pointing or convergent. Therefore, by the principle established above, there must exist at least one convergent interval within the entire scope. The previously detailed binary search iteratively reduces this scope in a way such that a convergent interval is always included, eventually narrowing the scope to a single convergent interval.

\subsection{Limitations}
While the monotonicity argument ensures that right-pointing and left-pointing intervals cannot directly switch roles when shifting in the direction they point (rightward for right-pointing, leftward for left-pointing), this guarantee does not hold when shifting in the opposite direction. Multiple local minima in Lloyd’s algorithm can produce patterns like:
\[
RRRCLLLRRRCLLL,
\]
where \(R\) denotes right-pointing, \(L\) denotes left-pointing, and \(C\) denotes convergent intervals. In such scenarios:
\begin{itemize}
    \item Moving a left-pointing interval to the right can produce a right-pointing interval (and vice versa).
    \item The algorithm still finds at least one convergent interval in \(O(\log n)\) time, but the found cluster boundary may correspond to a local minimum rather than the global optimum.
\end{itemize}

\subsection{Algorithm Guarantees}

By leveraging the monotonic behavior of the midpoint and the fact that a convergent interval must exist between the first right-pointing and last left-pointing intervals, the binary search efficiently identifies a convergent interval representing a cluster boundary. The algorithm achieves this in \(O(\log n)\) time without relying on iterative steps of Lloyd's algorithm.

In the presence of multiple local minima, the algorithm is guaranteed to converge to a valid solution, though it may not necessarily find the global optimum. Similar to the \(k\)-cluster algorithm, an initial preprocessing step is required, which includes sorting the data in \(O(n \log n)\) time and computing prefix sums in \(O(n)\) time.

\section{Summary}

In this chapter, we introduced two novel algorithms for solving the \(k\)-cluster problem in one-dimensional sorted data: the \textbf{\(k\)-cluster algorithm} and the \textbf{2-cluster algorithm}. Both methods leverage key observations about the structure of one-dimensional data to achieve significant computational efficiency.

\begin{itemize}
    \item The \textbf{\(k\)-cluster algorithm} combines greedy $k$-means++ initialization with efficient Lloyd’s iterations. By exploiting sorted data, prefix sums, and binary search:
        \begin{itemize}
            \item The initialization process achieves a complexity of \(O(l \cdot k^2 \cdot \log n)\), where \(k\) is the number of clusters and \(l\) is the number of local trials.
            \item Each iteration of Lloyd’s algorithm requires \(O(k \log n)\) time, ensuring efficient updates of cluster boundaries and centroids.
        \end{itemize}

    \item The \textbf{2-cluster algorithm} focuses on the specific case of \(k=2\), where the problem reduces to locating a single cluster boundary. Using a binary search over division intervals:
        \begin{itemize}
            \item The midpoint of centroids behaves monotonically, allowing us to refine the search scope iteratively.
            \item The algorithm achieves a total complexity of \(O(\log n)\) and guarantees convergence to a local minimum of Lloyd's algorithm.
        \end{itemize}
\end{itemize}

Both algorithms demonstrate how the structure of one-dimensional data enables faster computations compared to traditional clustering methods. In cases with multiple local minima, the solutions produced may not be globally optimal---a limitation shared with Lloyd's algorithm. Nonetheless, the proposed methods strike an effective balance between accuracy and computational efficiency, making them highly suitable for practical applications.

\chapter{Experiments}\label{chap:experiments}

This chapter presents experimental results comparing the proposed 1D $k$-means algorithms against the widely used \texttt{scikit-learn} implementation. The evaluation focuses on runtime performance and clustering quality, using both real and synthetic datasets.

\section{Implementation Details}

\sloppy
The proposed algorithms have been implemented as an open-source Python package, \texttt{flash1dkmeans}, available on GitHub\footnote{\url{https://github.com/SyphonArch/flash1dkmeans}} and PyPI\footnote{\url{https://pypi.org/project/flash1dkmeans/}}. The package is built on top of \texttt{NumPy} \cite{numpy} and \texttt{Numba} \cite{numba} for efficient computation. The implementation includes both the $k$-cluster and 2-cluster algorithms, as well as the preprocessing steps for the sorting and prefix sum computations. The preprocessing step is optional and can be disabled if the input data is already in adequate form. More details on the implementation can be found in the Appendix.

\section{Experimental Setup}

The experiments were conducted on two processors: an Intel i9-13900K and an Apple M3. For the clustering quality and runtime comparisons, only the i9-13900K results are presented, as trends were consistent across both processors. Both real and synthetic 1D datasets were used, with varying sizes and numbers of clusters. The real datasets include the \texttt{iris} and \texttt{california housing} datasets from \texttt{scikit-learn}, while synthetic datasets were generated using the \texttt{sklearn.datasets.make\_blobs} and \texttt{numpy.random.random\_sample} functions.

Comparisons were made against \texttt{scikit-learn}'s implementation of $k$-means\footnote{\url{https://scikit-learn.org/stable/modules/generated/sklearn.cluster.KMeans.html}} at default configuration, which utilizes greedy $k$-means++ initialization paired with Lloyd's algorithm. For fair comparison, only a single thread was used. The number of local trials for the greedy $k$-means++ initialization, $l$, defaults to $2 + \log k$ in \texttt{scikit-learn}. This was matched in \texttt{flash1dkmeans}.

\section{Results: Clustering Quality}\label{sec:clustering_quality}

\begin{figure}[H]
    \includegraphics[width=\textwidth]{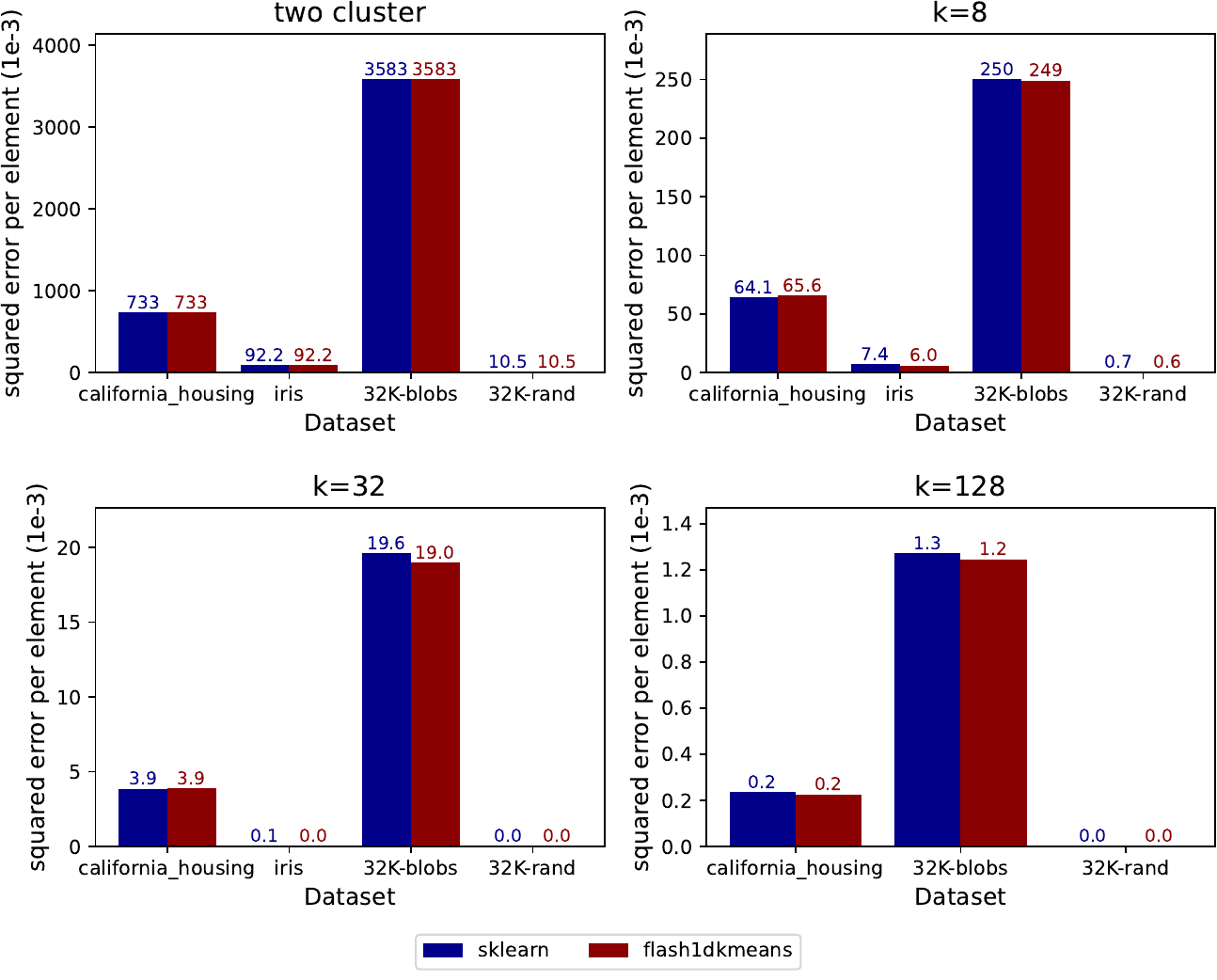}
    \caption{Squared error comparison of the proposed two-cluster algorithm and $k$-cluster algorithms in \texttt{flash1dkmeans} against \texttt{scikit-learn} on real and synthetic datasets. Lower is better.}
    \label{fig:inertia_comparison}
\end{figure}

The clustering quality comparison between the proposed 2-cluster algorithm and the $k$-cluster algorithms in \texttt{flash1dkmeans} and \texttt{scikit-learn} is shown in Figure~\ref{fig:inertia_comparison}. The \texttt{iris} dataset was excluded for $k=128$ as the number of data points was insufficient. Across the configurations, the clustering quality, measured by squared error, is consistent with the baseline $k$-means algorithm. For both the 2-cluster and $k$-cluster algorithms, this confirms that the proposed method produces clustering results equivalent or close to the $k$-means algorithm in \texttt{scikit-learn}.

\section{Results: Runtime Performance}\label{sec:runtime_performance}

\begin{figure}[H]
    \includegraphics[width=\textwidth]{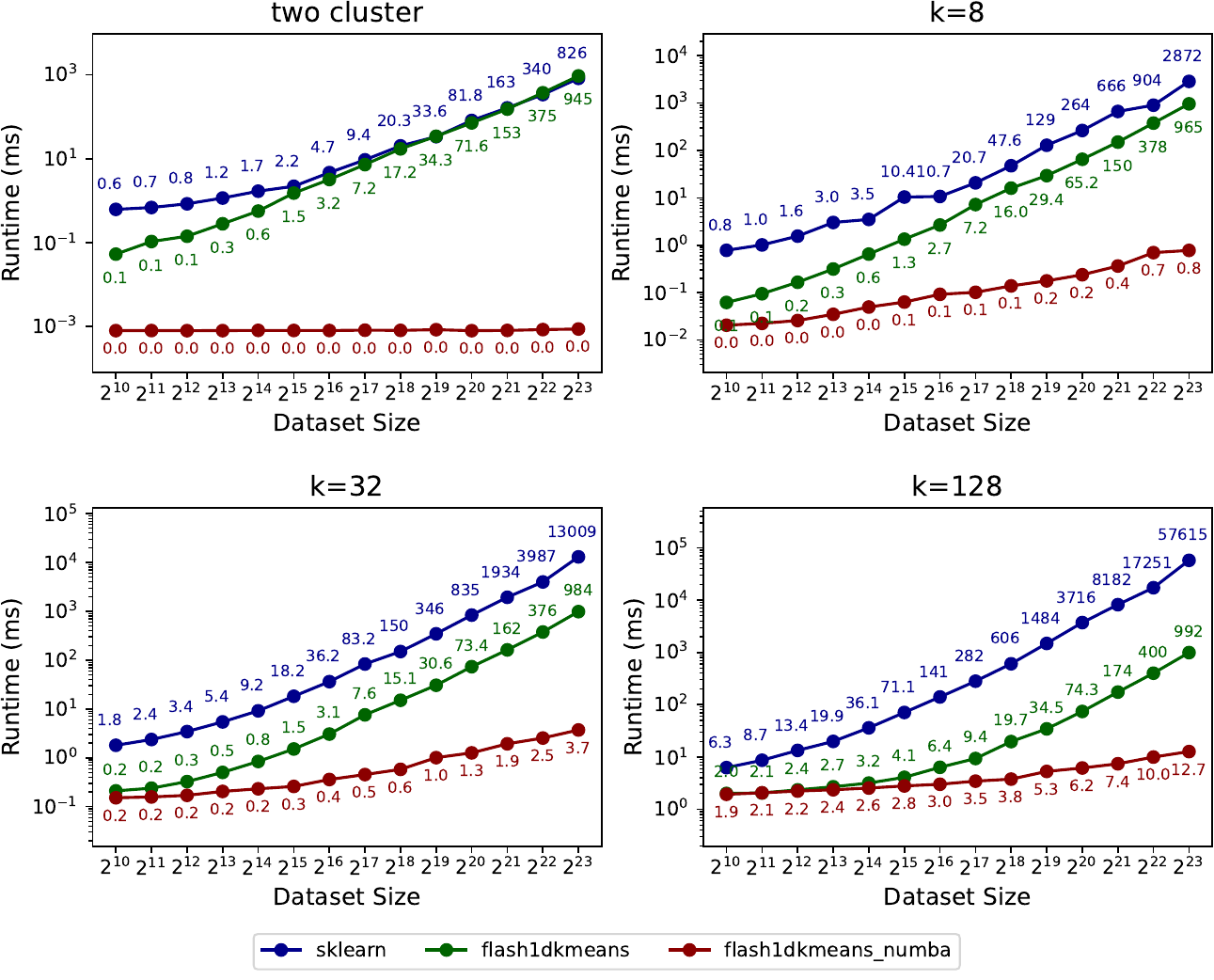}
    \caption{Runtime comparison of the proposed two-cluster algorithm and $k$-cluster algorithms in \texttt{flash1dkmeans} against \texttt{scikit-learn} on datasets of varying sizes. Lower is better.}
    \label{fig:runtime_comparison}
\end{figure}

Figure~\ref{fig:runtime_comparison} compares the runtime of the proposed 2-cluster and $k$-cluster algorithms in \texttt{flash1dkmeans} to \texttt{scikit-learn}. The runtime of \texttt{flash1dkmeans} includes preprocessing time, while \texttt{flash1dkmeans\_numba} measures only the main algorithm runtime, assuming preprocessed and sorted input data. 

The results confirm that the $k$-cluster algorithm in \texttt{flash1dkmeans} achieves substantial runtime improvements compared to \texttt{scikit-learn}, showcasing the logarithmic dependence on dataset size. For example, for \(k=128\) and dataset size \(2^{23}\), the speedup exceeds 4500x. However, for the 2-cluster algorithm on larger datasets, the \(O(n \log n)\) preprocessing time becomes a bottleneck, limiting the overall speedup. Nevertheless, the main algorithm runtimes confirm the log-time efficiency of the proposed optimizations, and \texttt{flash1dkmeans} with the $O(n \log n)$ preprocessing still outperforms \texttt{scikit-learn} by a large margin in most cases.

\section{Results: LLM Quantization}\label{sec:llm_quantization}

\begin{table}[H]
    \centering
    \resizebox{\linewidth}{!}{%
    \begin{tabular}{|l|c c c|c c c|}
        \cline{2-7}
        \multicolumn{1}{c|}{} & \multicolumn{3}{c|}{\textbf{Seed (k-cluster algorithm)}} & \multicolumn{3}{c|}{\textbf{Upscale (2-cluster algorithm)}} \\
        \hline
        \textbf{Processor} & \textbf{sklearn} & \textbf{flash1dkmeans} & \textbf{speedup} & \textbf{sklearn} & \textbf{flash1dkmeans} & \textbf{speedup} \\
        \hline
        \textbf{i9-13900K} & 433 ms & 56 ms & 7.7x & 10551 ms & 34 ms & 310x \\
        \textbf{Apple M3}  & 572 ms & 114 ms & 5.0x & 7585 ms  & 29 us & 262x \\
        \hline
    \end{tabular}
    }
    \caption{Performance comparison of \texttt{flash1dkmeans} against \texttt{scikit-learn} for both $k$-cluster and 2-cluster algorithms on i9-13900K and Apple M3 processors.}
    \label{tab:anyprec}
\end{table}

Table~\ref{tab:anyprec} compares the performance of \texttt{flash1dkmeans} and \texttt{scikit-learn} in the two-step LLM quantization process proposed in Any-Precision LLM \cite{anyprec}. As an example, we quantize the 214th output channel (size 14,336) of the \texttt{mlp.down\_proj} linear module in the 8th layer of Llama-3-8B \cite{llama3} from 3-bit to 8-bit, as described in the original work. We measure the total time over 100 runs for reliability. 

The first step involves generating a seed model using $k$-means clustering, followed by incrementally subdividing clusters into two via $k$-means in the second step. The preprocessing from the seed generation step is reusable during the upscale phase, enabling the $O(\log n)$-time 2-cluster algorithm to be particularly efficient. Results demonstrate that \texttt{flash1dkmeans} achieves remarkable speedups, surpassing 300x in the upscale step on an i9-13900K processor.

\section{Summary}

Experimental results demonstrate that the proposed algorithms in \texttt{flash1dkmeans} achieve comparable clustering quality to \texttt{scikit-learn} while offering significant runtime improvements. The $k$-cluster algorithm exhibits logarithmic time complexity with speedups exceeding 4500x on large datasets, while the 2-cluster algorithm achieves remarkable efficiency when given preprocessed inputs. These results validate the effectiveness of the proposed optimizations for both general clustering tasks and emerging applications such as LLM quantization.

\chapter{Conclusion}\label{chap:conclusion}

This thesis presents optimized algorithms for $k$-means++ initialization and Lloyd’s algorithm, specifically designed for one-dimensional (1D) clustering. By leveraging the structure of sorted data, the proposed methods replace the linear dependence on dataset size \(n\) with logarithmic factors, significantly improving computational efficiency while maintaining clustering quality. The optimized greedy $k$-means++ initialization achieves a time complexity of \(O(l \cdot k^2 \cdot \log n)\), while Lloyd’s algorithm iterations achieve \(O(i \cdot k \cdot \log n)\), where \(l\) is the number of local trials, and \(i\) the number of iterations. Additionally, the binary search-based algorithm for the two-cluster case deterministically converges to a Lloyd’s algorithm solution in \(O(\log n)\), bypassing iterative refinements entirely.

These methods offer significant benefits for latency-critical tasks and large-scale clustering problems. The practical relevance of these contributions is demonstrated in key applications such as quantization for large language models (LLMs), where the optimized algorithms achieve a 300-fold speedup over \texttt{scikit-learn}, as discussed in Section~\ref{sec:llm_quantization}. To ensure ease of use, a Python implementation of the proposed methods, optimized with just-in-time (JIT) compilation, is provided for practical deployment.

By exploiting the structure of 1D data, this thesis demonstrates that substantial computational improvements are achievable for $k$-means clustering, reducing dataset size dependencies from linear to logarithmic. These contributions address an important gap in the existing literature and deliver efficient, high-performance clustering tools ready for real-world applications.
\cleardoublepage

\pagenumbering{roman}
\setcounter{page}{1} 
\renewcommand{\bibname}{References}
\phantomsection 
\addcontentsline{toc}{chapter}{References} 
\bibliographystyle{IEEEtran}
\bibliography{bib}

\chapter*{Appendix}\label{chap:appendix}
\addcontentsline{toc}{chapter}{Appendix}

\renewcommand{\thesection}{\Alph{section}.} 
\setcounter{section}{0} 

\section{$k$-Cluster Algorithm Implementation}
\label{sec:kclustercode}

Provided below is the Python 3 implementation of the $k$-cluster algorithm discussed in this work. The \texttt{Numba} and \texttt{Numpy} packages are required, as well as the definition of macros like \texttt{ARRAY\_INDEX\_DTYPE}. For the fully integrated library please refer to section \ref{sec:github}.

\begin{lstlisting}[language=Python]
@numba.njit(cache=True)
def numba_kmeans_1d_k_cluster(
        sorted_X,
        n_clusters,
        max_iter,
        weights_prefix_sum, weighted_X_prefix_sum,
        weighted_X_squared_prefix_sum,
        start_idx,
        stop_idx,
        random_state=None,
):
    """An optimized kmeans for 1D data with n clusters.
    Exploits the fact that the data is 1D to optimize the calculations.
    Time complexity: O(k ^ 2 * log(k) * log(n) + i * log(n) * k)

    Args:
        sorted_X: np.ndarray
            The input data. Should be sorted in ascending order.
        n_clusters: int
            The number of clusters to generate
        max_iter: int
            The maximum number of iterations to run
        weights_prefix_sum: np.ndarray
            The prefix sum of the weights. Should be None if the data is unweighted.
        weighted_X_prefix_sum: np.ndarray
            The prefix sum of the weighted X
        weighted_X_squared_prefix_sum: np.ndarray
            The prefix sum of the weighted X squared
        start_idx: int
            The start index of the range to consider
        stop_idx: int
            The stop index of the range to consider
        random_state: int or None
            The random seed to use.

    Returns:
        centroids: np.ndarray
            The centroids of the clusters
        cluster_borders: np.ndarray
            The borders of the clusters
    """
    # set random_state
    set_np_seed_njit(random_state)

    cluster_borders = np.empty(n_clusters + 1, dtype=ARRAY_INDEX_DTYPE)
    cluster_borders[0] = start_idx
    cluster_borders[-1] = stop_idx

    centroids = _kmeans_plusplus(
        sorted_X, n_clusters,
        weights_prefix_sum, weighted_X_prefix_sum,
        weighted_X_squared_prefix_sum,
        start_idx, stop_idx,
    )
    sorted_centroids = np.sort(centroids)

    for _ in range(max_iter):
        new_cluster_borders = _centroids_to_cluster_borders(sorted_X, sorted_centroids, start_idx, stop_idx)

        if np.array_equal(cluster_borders, new_cluster_borders):
            break

        cluster_borders[:] = new_cluster_borders
        for i in range(n_clusters):
            cluster_start = cluster_borders[i]
            cluster_end = cluster_borders[i + 1]

            if cluster_end < cluster_start:
                raise ValueError("Cluster end is less than cluster start")

            if cluster_start == cluster_end:
                continue

            cluster_weighted_X_sum = query_prefix_sum(weighted_X_prefix_sum, cluster_start, cluster_end)
            cluster_weight_sum = query_prefix_sum(weights_prefix_sum, cluster_start, cluster_end)

            if cluster_weight_sum == 0:
                # if the sum of the weights is zero, we set the centroid to the mean of the cluster
                sorted_centroids[i] = sorted_X[cluster_start:cluster_end].mean()
            else:
                sorted_centroids[i] = cluster_weighted_X_sum / cluster_weight_sum

    return sorted_centroids, cluster_borders


@numba.njit(cache=True)
def _rand_choice_prefix_sum(arr, prob_prefix_sum, start_idx, stop_idx):
    """Randomly choose an element from arr according to the probability distribution given by prob_prefix_sum
    
    Time complexity: O(log n)

    Args:
        arr: np.ndarray
            The array to choose from
        prob_prefix_sum: np.ndarray
            The prefix sum of the probability distribution
        start_idx: int
            The start index of the range to consider
        stop_idx: int
            The stop index of the range to consider

    Returns:
        The chosen element
    """
    total_prob = query_prefix_sum(prob_prefix_sum, start_idx, stop_idx)
    selector = np.random.random_sample() * total_prob

    # Because we are using start_idx as the base, but the prefix sum is calculated from 0,
    # we need to adjust the selector if start_idx is not 0.
    adjusted_selector = selector + prob_prefix_sum[start_idx - 1] if start_idx > 0 else selector

    # Search for the index of the selector in the prefix sum, and add start_idx to get the index in the original array
    idx = np.searchsorted(prob_prefix_sum[start_idx:stop_idx], adjusted_selector) + start_idx

    return arr[idx]


@numba.njit(cache=True)
def _centroids_to_cluster_borders(X, sorted_centroids, start_idx, stop_idx):
    """Converts the centroids to cluster borders.
    The cluster borders are where the clusters are divided.
    The centroids must be sorted.

    Time complexity: O(k * log n)

    Args:
        X: np.ndarray
            The input data. Should be sorted in ascending order.
        sorted_centroids: np.ndarray
            The sorted centroids
        start_idx: int
            The start index of the range to consider
        stop_idx: int
            The stop index of the range to consider

    Returns:
        np.ndarray: The cluster borders
    """
    midpoints = (sorted_centroids[:-1] + sorted_centroids[1:]) / 2
    cluster_borders = np.empty(len(sorted_centroids) + 1, dtype=ARRAY_INDEX_DTYPE)
    cluster_borders[0] = start_idx
    cluster_borders[-1] = stop_idx
    cluster_borders[1:-1] = np.searchsorted(X[start_idx:stop_idx], midpoints) + start_idx
    return cluster_borders


@numba.njit(cache=True)
def _calculate_inertia(sorted_centroids, centroid_ranges,
                       weights_prefix_sum, weighted_X_prefix_sum, weighted_X_squared_prefix_sum,
                       stop_idx):
    """Calculates the inertia of the clusters given the centroids.
    The inertia is the sum of the squared distances of each sample to the closest centroid.
    The calculations are done efficiently using prefix sums.

    Time complexity: O(k)

    Args:
        sorted_centroids: np.ndarray
            The centroids of the clusters
        centroid_ranges: np.ndarray
            The borders of the clusters
        weights_prefix_sum: np.ndarray
            The prefix sum of the weights. Should be None if the data is unweighted.
        weighted_X_prefix_sum: np.ndarray
            The prefix sum of the weighted X
        weighted_X_squared_prefix_sum: np.ndarray
            The prefix sum of the weighted X squared
        stop_idx: int
            The stop index of the range to consider
    """
    # inertia = sigma_i(w_i * abs(x_i - c)^2) = sigma_i(w_i * (x_i^2 - 2 * x_i * c + c^2))
    #         = sigma_i(w_i * x_i^2) - 2 * c * sigma_i(w_i * x_i) + c^2 * sigma_i(w_i)
    #         = sigma_i(weighted_X_squared) - 2 * c * sigma_i(weighted_X) + c^2 * sigma_i(weight)
    #  Note that the centroid c is the CLOSEST centroid to x_i, so the above calculation must be done for each cluster

    inertia = 0
    for i in range(len(sorted_centroids)):
        start = centroid_ranges[i]
        end = centroid_ranges[i + 1]

        if start >= stop_idx:
            break
        if end >= stop_idx:
            end = stop_idx

        if start == end:
            continue

        cluster_weighted_X_squared_sum = query_prefix_sum(weighted_X_squared_prefix_sum, start, end)
        cluster_weighted_X_sum = query_prefix_sum(weighted_X_prefix_sum, start, end)
        cluster_weight_sum = query_prefix_sum(weights_prefix_sum, start, end)

        inertia += (cluster_weighted_X_squared_sum - 2 * sorted_centroids[i] * cluster_weighted_X_sum +
                    sorted_centroids[i] ** 2 * cluster_weight_sum)

    return inertia


@numba.njit(cache=True)
def _rand_choice_centroids(X, centroids,
                           weights_prefix_sum, weighted_X_prefix_sum, weighted_X_squared_prefix_sum,
                           sample_size, start_idx, stop_idx):
    """Randomly choose sample_size elements from X, weighted by the distance to the closest centroid.
    The weighted logic is implemented efficiently by utilizing the _calculate_inertia function.

    Time complexity: O(l * k * log n)

    Args:
        X: np.ndarray
            The input data. Should be sorted in ascending order.
        centroids: np.ndarray
            The centroids of the clusters
        is_weighted: bool
            Whether the data is weighted. If True, the weighted versions of the arrays should be provided.
        weights_prefix_sum: np.ndarray
            The prefix sum of the weights. Should be None if the data is unweighted.
        weighted_X_prefix_sum: np.ndarray
            The prefix sum of the weighted X
        weighted_X_squared_prefix_sum: np.ndarray
            The prefix sum of the weighted X squared
        sample_size: int
            The number of samples to choose
        start_idx: int
            The start index of the range to consider
        stop_idx: int
            The stop index of the range to consider

    Returns:
        np.ndarray: The chosen samples
    """
    sorted_centroids = np.sort(centroids)  # O(k log k)
    cluster_borders = _centroids_to_cluster_borders(X, sorted_centroids, start_idx, stop_idx)  # O(k log n)
    total_inertia = _calculate_inertia(sorted_centroids, cluster_borders,  # O(k)
                                       weights_prefix_sum, weighted_X_prefix_sum,
                                       weighted_X_squared_prefix_sum, stop_idx)
    selectors = np.random.random_sample(sample_size) * total_inertia
    results = np.empty(sample_size, dtype=centroids.dtype)

    for i in range(sample_size):  # O(l k log n)
        selector = selectors[i]
        floor = start_idx + 1
        ceiling = stop_idx
        while floor < ceiling:
            stop_idx_cand = (floor + ceiling) // 2
            inertia = _calculate_inertia(sorted_centroids, cluster_borders,  # O(k)
                                         weights_prefix_sum, weighted_X_prefix_sum,
                                         weighted_X_squared_prefix_sum, stop_idx_cand)
            if inertia < selector:
                floor = stop_idx_cand + 1
            else:
                ceiling = stop_idx_cand
        results[i] = X[floor - 1]

    return results


@numba.njit(cache=True)
def _kmeans_plusplus(X, n_clusters,
                     weights_prefix_sum, weighted_X_prefix_sum, weighted_X_squared_prefix_sum,
                     start_idx, stop_idx):
    """An optimized version of the kmeans++ initialization algorithm for 1D data.
    The algorithm is optimized for 1D data and utilizes prefix sums for efficient calculations.

    Time complexity: = O(k ^ 2 * log k * log n)

    Args:
        X: np.ndarray
            The input data
        n_clusters: int
            The number of clusters to choose
        weights_prefix_sum: np.ndarray
            The prefix sum of the weights. Should be None if the data is unweighted.
        weighted_X_prefix_sum: np.ndarray
            The prefix sum of the weighted X
        weighted_X_squared_prefix_sum: np.ndarray
            The prefix sum of the weighted X squared

    Returns:
        np.ndarray: The chosen centroids
    """
    centroids = np.empty(n_clusters, dtype=X.dtype)
    n_local_trials = 2 + int(np.log(n_clusters))

    # First centroid is chosen randomly according to sample_weight
    centroids[0] = _rand_choice_prefix_sum(X, weights_prefix_sum, start_idx, stop_idx)  # O(log n)

    for c_id in range(1, n_clusters):  # O(k^2 l log n)
        # Choose the next centroid randomly according to the weighted distances
        # Sample n_local_trials candidates and choose the best one

        centroid_candidates = _rand_choice_centroids(  # O(l k log n)
            X, centroids[:c_id],
            weights_prefix_sum, weighted_X_prefix_sum,
            weighted_X_squared_prefix_sum, n_local_trials,
            start_idx, stop_idx
        )

        best_inertia = np.inf
        best_centroid = None
        for i in range(len(centroid_candidates)): # O(l k log n)
            # O(k log k + k log n + k) = O(k log n), as k <= n
            centroids[c_id] = centroid_candidates[i]
            sorted_centroids = np.sort(centroids[:c_id + 1]) # O(k log k), I think we could avoid centroid sorting and use some linear algorithm, but the gain would be minimal, especially considering that k <= n, and most times k << n
            centroid_ranges = _centroids_to_cluster_borders(X, sorted_centroids, start_idx, stop_idx)  # O(k log n)
            inertia = _calculate_inertia(sorted_centroids, centroid_ranges,  # O(k)
                                         weights_prefix_sum, weighted_X_prefix_sum,
                                         weighted_X_squared_prefix_sum, stop_idx)
            if inertia < best_inertia:
                best_inertia = inertia
                best_centroid = centroid_candidates[i]
        centroids[c_id] = best_centroid

    return centroids
\end{lstlisting}

\section{2-Cluster Agorithm Implementation}
\label{sec:2clustercode}

Provided below is the Python 3 implementation of the $2$-cluster algorithm discussed in this work. The \texttt{Numba} and \texttt{Numpy} packages are required, as well as the definition of macros like \texttt{ARRAY\_INDEX\_DTYPE}. For the fully integrated library please refer to section \ref{sec:github}.

\begin{lstlisting}[language=Python]
@numba.njit(cache=True)
def numba_kmeans_1d_two_cluster(
        sorted_X,
        weights_prefix_sum,
        weighted_X_prefix_sum,
        start_idx,
        stop_idx,
):
    """An optimized kmeans for 1D data with 2 clusters, weighted version.
    Utilizes a binary search to find the optimal division point.
    Time complexity: O(log(n))

    Args:
        sorted_X: np.ndarray
            The input data. Should be sorted in ascending order.
        weights_prefix_sum: np.ndarray
            The prefix sum of the sample weights. Should be None if the data is unweighted.
        weighted_X_prefix_sum: np.ndarray
            The prefix sum of (the weighted) X.
        start_idx: int
            The start index of the range to consider.
        stop_idx: int
            The stop index of the range to consider.

    Returns:
        centroids: np.ndarray
            The centroids of the two clusters, shape (2,)
        cluster_borders: np.ndarray
            The borders of the two clusters, shape (3,)

    WARNING: X should be sorted in ascending order before calling this function.
    """
    size = stop_idx - start_idx
    centroids = np.empty(2, dtype=sorted_X.dtype)
    cluster_borders = np.empty(3, dtype=ARRAY_INDEX_DTYPE)
    cluster_borders[0] = start_idx
    cluster_borders[2] = stop_idx
    # Remember to set cluster_borders[1] as the division point

    if size == 1:
        centroids[0], centroids[1] = sorted_X[start_idx], sorted_X[start_idx]
        cluster_borders[1] = start_idx + 1
        return centroids, cluster_borders

    if size == 2:
        centroids[0], centroids[1] = sorted_X[start_idx], sorted_X[start_idx + 1]
        cluster_borders[1] = start_idx + 1
        return centroids, cluster_borders

    # Now we know that there are at least 3 elements

    # If the sum of the sample weight in the range is 0, we assume that the data is unweighted
    if query_prefix_sum(weights_prefix_sum, start_idx, stop_idx) == 0:
        # We need to recalculate the prefix sum, as previously it would have been all zeros
        X_casted = sorted_X.astype(PREFIX_SUM_DTYPE)
        X_prefix_sum = np.cumsum(X_casted)
        return numba_kmeans_1d_two_cluster_unweighted(sorted_X, X_prefix_sum, start_idx, stop_idx)
    else:
        # Check if there is only one nonzero sample weight
        total_weight = query_prefix_sum(weights_prefix_sum, start_idx, stop_idx)
        sample_weight_prefix_sum_within_range = weights_prefix_sum[start_idx:stop_idx]
        final_increase_idx = np.searchsorted(
            sample_weight_prefix_sum_within_range,
            sample_weight_prefix_sum_within_range[-1]
        )
        final_increase_amount = query_prefix_sum(weights_prefix_sum,
                                                 start_idx + final_increase_idx,
                                                 start_idx + final_increase_idx + 1)
        if total_weight == final_increase_amount:
            # If there is only one nonzero sample weight, we need to return the corresponding weight as the centroid
            # and set all elements to the left cluster
            nonzero_weight_index = start_idx + final_increase_idx
            centroids[0], centroids[1] = sorted_X[nonzero_weight_index], sorted_X[nonzero_weight_index]
            cluster_borders[1] = stop_idx
            return centroids, cluster_borders

    # Now we know that there are at least 3 elements and at least 2 nonzero weights

    # KMeans with 2 clusters on 1D data is equivalent to finding a division point.
    # The division point can be found by doing a binary search on the prefix sum.

    # We will do a search for the division point,
    # where we search for the optimum number of elements in the first cluster
    # We don't want empty clusters, so we set the floor and ceiling to start_idx + 1 and stop_idx - 1
    floor = start_idx + 1
    ceiling = stop_idx - 1
    left_centroid = None
    right_centroid = None

    while floor < ceiling:
        division_point = (floor + ceiling) // 2
        # If the left cluster has no weight, we need to move the floor up
        left_weight_sum = query_prefix_sum(weights_prefix_sum, start_idx, division_point)
        if left_weight_sum == 0:
            floor = division_point + 1
            continue
        right_weight_sum = query_prefix_sum(weights_prefix_sum, division_point, stop_idx)
        # If the right cluster has no weight, we need to move the ceiling down
        if right_weight_sum == 0:
            ceiling = division_point - 1
            continue

        left_centroid = query_prefix_sum(weighted_X_prefix_sum, start_idx, division_point) / left_weight_sum
        right_centroid = query_prefix_sum(weighted_X_prefix_sum, division_point, stop_idx) / right_weight_sum

        new_division_point_value = (left_centroid + right_centroid) / 2
        if sorted_X[division_point - 1] <= new_division_point_value:
            if new_division_point_value <= sorted_X[division_point]:
                # The new division point matches the previous one, so we can stop
                break
            else:
                floor = division_point + 1
        else:
            ceiling = division_point - 1

    # recalculate division point based on final floor and ceiling
    division_point = (floor + ceiling) // 2

    # initialize variables in case the loop above does not run through
    if left_centroid is None:
        left_centroid = (query_prefix_sum(weighted_X_prefix_sum, start_idx, division_point) /
                         query_prefix_sum(weights_prefix_sum, start_idx, division_point))
    if right_centroid is None:
        right_centroid = (query_prefix_sum(weighted_X_prefix_sum, division_point, stop_idx) /
                          query_prefix_sum(weights_prefix_sum, division_point, stop_idx))

    # avoid using lists to allow numba.njit
    centroids[0] = left_centroid
    centroids[1] = right_centroid

    cluster_borders[1] = division_point
    return centroids, cluster_borders
\end{lstlisting}

\section{Library Implementation}
\label{sec:github}
The algorithms detailed in this thesis have been published open-source as \texttt{flash1dkmeans}.

\noindent Github respository: \url{https://github.com/SyphonArch/flash1dkmeans} \newline
\texttt{PyPI}: \url{https://pypi.org/project/flash1dkmeans/}
\cleardoublepage

\chapter*{국문초록}
\addcontentsline{toc}{chapter}{국문초록}

클러스터링은 머신러닝에서 핵심적인 과제로, $k$-means는 단순성과 효율성 덕분에 널리 사용되는 알고리즘이다. 1차원(1D) 클러스터링은 많은 실제 응용에서 발생하지만, 기존 $k$-means 구현체들은 1D 데이터의 구조를 효과적으로 활용하지 못해 비효율이 존재한다. 본 논문에서는 정렬된 데이터, 누적합 배열, 이진탐색의 특성을 활용하여 1D 클러스터링에 최적화된 $k$-means++ 초기화 및 Lloyd 알고리즘을 제안한다.

본 논문은 다음과 같은 로그 시간 알고리즘을 제시한다: (1) \(k\)-cluster 알고리즘은 greedy $k$-means++ 초기화에서 \(O(l \cdot k^2 \cdot \log n)\) 시간복잡도, Lloyd 알고리즘에서 \(O(i \cdot k \cdot \log n)\) 시간복잡도를 달성한다. 여기서 \(n\)은 데이터셋 크기, \(k\)는 클러스터 개수, \(l\)은 greedy $k$-means++ local trials 수, \(i\)는 Lloyd 알고리즘 반복 횟수를 나타낸다. (2) 2-cluster 알고리즘은 이진탐색을 활용하여 \(O(\log n)\) 시간복잡도로 작동하며, 반복 없이 Lloyd 알고리즘의 국소 최적해에 빠르게 수렴한다.

벤치마크 결과, 제시된 알고리즘은 대규모 데이터셋에서 \texttt{scikit-learn} 대비 4500배 이상의 속도 향상을 달성하면서도 within-cluster sum of squares (WCSS) 품질을 유지한다. 또한, 대규모 언어 모델(LLMs) 양자화와 같은 최신 응용에서도 300배 이상의 속도 향상을 보여준다.

본 연구는 1D $k$-means clustering의 이론과 실제 간 간극을 좁히는 효율적이고 실용적인 알고리즘을 제안한다. 제시된 알고리즘은 JIT 컴파일을 통해 최적화된 오픈소스 Python 라이브러리로 구현되었으며, 실제 응용에 쉽게 통합할 수 있도록 설계되었다.

\vspace*{1cm}
\textbf{주요어: $k$-means 클러스터링, Lloyd 알고리즘, $k$-means++ 초기화, 일차원 클러스터링, 이진탐색, 누적합}

\end{document}